\documentclass[aps,final,notitlepage,oneside,twocolumn,nobibnotes,nofootinbib,
superscriptaddress,noshowpacs,centertags]{revtex4-1}
\pdfoutput=1

\usepackage[utf8]{inputenc}
\usepackage[english]{babel}
\usepackage{graphicx}
\usepackage[dvipsnames]{xcolor}
\usepackage[colorlinks,urlcolor=Black,citecolor=Black,linkcolor=Black,pdfusetitle,breaklinks]{hyperref}
\hypersetup{breaklinks=true,colorlinks=true}
\usepackage{latexsym}
\usepackage{amssymb}
\usepackage{amsmath}
\usepackage{float}
\usepackage{color}
\usepackage{soul} 
\usepackage{natbib}

\newcommand{\eq}{{\rm eq}}
\newcommand{\np}{{\rm np}}

\renewcommand{\c}{{\rm c}}
\newcommand{\p}{{\rm p}}

\newcommand{\Msun}{M_\odot}
\newcommand{\fbh}{f_{\rm PBH}}
\newcommand{\Mch}{M_{\rm ch}}

\begin{document}

\title{Redshift evolution of primordial black hole merger rate}

\author{Viktor Stasenko}\thanks{e-mail: vdstasenko@mephi.ru}
\affiliation{National Research Nuclear University MEPhI, Moscow 115409, Russia}
\affiliation{Novosibirsk State University, Novosibirsk 630090, Russia}

\begin{abstract}
The gravitational wave signals detected by the LIGO-Virgo-KAGRA collaboration can be explained by mergers of binary primordial black holes (PBHs) formed in the radiation dominated epoch. However, in early structures induced by the Poisson distribution of PBHs, a significant fraction of binaries are perturbed and avoid mergers. In addition, the internal dynamics of early halos lead to the formation of dense primordial black hole clusters within a few Hubble times from the moment of halo formation. In such clusters PBH binaries are effectively formed and their mergers potentially dominate in the modern era. We obtained that the PBH merger rate changes with redshift as $\mathcal{R} \propto (1 + z)^\beta$, where $\beta = 1.4 - 2.2$ reflects the influence of PBH clustering and depends on both $z$ and $f_{\rm PBH}$. The observed merger rate constraints the fraction of PBHs of tens solar masses in the composition of dark matter $f_{\rm PBH} \lesssim 0.001 - 0.1$ in dependence of the clustering efficiency. 

\end{abstract}

\maketitle 

\section{Introduction}

The collapse of hypothetical regions with high overdensity leads to the formation of primordial black holes (PBHs) in the early Universe~\cite{1971MNRAS.152...75H, 1974MNRAS.168..399C}. The most popular mechanism is that these fluctuations arise during cosmic inflation~\cite{PhysRevD.47.4244, PhysRevD.50.7173, Garcia-Bellido:1996mdl, Kohri:2007qn, Clesse:2015wea, Garcia-Bellido:2017mdw}. These black, depending on their mass, could make up some fraction (or even all) of the dark matter (DM)~\cite{Carr:2020gox, Carr:2020xqk,Green:2020jor}. In particular, the discovery of gravitational waves by the LIGO-Virgo collaboration~\cite{LIGOScientific:2016aoc, LIGOScientific:2016sjg} has become the reason for active interest in PBHs with masses $m = 10 - 100\, \Msun$~\cite{Bird:2016dcv,
Sasaki:2016jop, Clesse:2016ajp, Clesse:2016vqa, Blinnikov:2016bxu, Cholis:2016kqi, Wang:2016ana, Raidal:2017mfl, 2017PhRvD..96l3523A, Kavanagh:2018ggo, Raidal:2018bbj, Chen:2018czv, Liu:2018ess, DeLuca:2020qqa,Jedamzik:2020ypm,Jedamzik:2020omx,Hutsi:2020sol}. 

One of the possibilities for the formation of PBH binaries is the decoupling of a PBH pair from the Hubble flow in the radiation
dominated Universe~\cite{Nakamura:1997sm,Ioka:1998nz,Sasaki:2016jop}. Henceforth such binaries will be called early binaries. After formation, the binary gradually shrinks due to the emission of gravitational waves and eventually merges. The Poisson initial distribution of PBHs also leads to the active formation of early dark structures if the fraction of PBHs in DM is $\fbh \gtrsim 0.01$~\cite{Afshordi:2003zb, Carr:2018rid, Hutsi:2019hlw, Inman:2019wvr, Murgia:2019duy}. Early binaries are perturbed in early halos during interactions with other PBHs and therefore avoid mergers. To satisfy the observed black holes merger rate by LIGO-Virgo-KAGRA (LVK) collaboration, the fraction of PBHs with masses $\sim 10 \, \Msun$ should be $\fbh \lesssim 0.001 - 0.1$~\cite{Raidal:2018bbj, Vaskonen:2019jpv, Stasenko:2023zmf}. The lower limit corresponds to the absence of clustering. The upper limit in turn implies that a significant fraction of binaries are perturbed in clusters and avoid mergers. In the last case the PBH merger rate is significantly reduced.

PBH binaries are also formed in the DM halo due to the emission of energy in the form of gravitational waves during pair scattering with a small impact parameter~\cite{Bird:2016dcv, Clesse:2016vqa}. We will call them late binaries. This mechanism is thought to make a subdominant contribution to the PBH merger rate~\cite{2017PhRvD..96l3523A}, but in the present work we show that this is not always the case. As noted earlier, the interaction of PBHs in the Poisson DM halos leads to a decrease in the merger rate of early binaries. At the same time, these gravitational interactions lead to the formation of dense PBH clusters in early halos~\cite{Stasenko:2023zmf}. These clusters are a suitable place for the effective formation of late binaries. 

The observation of gravitational waves by the LVK collaboration is not enough to answer the question about the nature of merging black holes. One of the distinctive signatures for PBHs is mergers at high redshifts  $z \gtrsim 30$~\cite{Chen:2019irf, Franciolini:2023opt, Branchesi:2023mws, Ng:2022agi, Ng:2022vbz, LISACosmologyWorkingGroup:2023njw}, since at this era it is difficult to produce mergers of black holes of astrophysical origin \cite{Koushiappas:2017kqm}. On the other hand, for PBH binaries forming in the early Universe, the merger rate increases monotonically with redshift $\mathcal{R} \propto (t(z) / t_0)^{-34/37}$~\cite{DeLuca:2020qqa, Vaskonen:2019jpv}, here $t_0$ is the age of the Universe. In this work we show that this dependence is valid only in the case of $\fbh < 10^{-3}$. For large PBH contributions to the DM composition, clustering effects cause the merger rate to increase faster with redshift. In addition, at low redshifts $z \lesssim 5$ mergers of late binaries in PBH clusters potentially predominate over early ones. Future ground-based third-generation gravitational wave detectors Einstein Telescope~\cite{Punturo:2010zz, Maggiore:2019uih} and Cosmic Explorer~\cite{Reitze:2019iox} will be able to observe black hole mergers of tens solar masses at redshifts up to $ z \sim 100$~\cite{Ng:2022agi, Franciolini:2023opt, Branchesi:2023mws}. Future observations will make it possible to reconstruct the time evolution of the merger rate and answer the question about the contribution of PBHs to the composition of dark matter. 

In this work, we consider two-component dark matter, consisting of primordial black holes with mass $m = 30 \, \Msun$ and light unknown particles. Low-mass PBHs can also act as dark matter particles~\cite{Carr:2019kxo, DeLuca:2020ioi, DeLuca:2020agl}. This work takes into account the gravitational interaction of PBHs with DM particles. In our analysis we mainly focus on the situation that the fraction of $30 \, \Msun$ PBHs in the dark matter composition is $\fbh \approx 0.01 - 0.1$. We also assume that PBHs are not clustered in space at the moment of their formation above the Poisson distribution~\cite{Rubin:2001yw, Belotsky:2018wph, Desjacques:2018wuu, Young:2019gfc, Ding:2019tjk, Atal:2020igj, Eroshenko:2023bbe}. Therefore, binaries are formed according to the scenario of~\cite{Nakamura:1997sm, Sasaki:2016jop}. The paper is organized as follows. In Sec.~\ref{sec2} we consider the internal evolution of early DM halos formed due to Poisson noise of PBHs. We show that the evolution of early structures naturally leads to the formation of dense primordial black hole clusters. In Sec.~\ref{sec3} we estimate the survival of clusters during the structures formation. Next, Sec.~\ref{sec4} examines the formation of PBH binaries in both the early and late Universe and estimates the merger rate of late binaries in clusters. Section.~\ref{sec5} studies the evolution of the PBH merger rate with cosmic time. Finally, we discuss our results in Sec.~\ref{sec6}.

\section{Dynamics of the early dark matter halo} \label{sec2}

Poisson noise in the initial spatial distribution of PBHs induces early formation of structures~\cite{Afshordi:2003zb, Inman:2019wvr}. The magnitude of fluctuations in the volume containing $N$ PBHs is estimated as $\delta_{\rm p} \sim \fbh / \sqrt{N}$~\cite{Carr:2018rid}. The variance of fluctuations on the mass scale $M$ is
\begin{equation}
    S_\p(M,z)  = \frac{m \fbh}{M} D^2(z),
\end{equation}
here $D(z)$ is the growth factor for isocurvature perturbations \cite{2011itec.book.....G}
\begin{equation}
    D(z) \approx \frac{3}{2} \left ( \frac{1 + z_{\rm eq}}{1 + z} \right),
    \label{Dz}
\end{equation}
where $z_{\eq} \approx 3400$ is the redshift of matter-radiation equality. Here we neglect standard adiabatic inflationary fluctuations, because at redshifts $z > 10$ their contribution is much less than Poisson noise. However, in the modern era, their contribution is important. We will consider the impact of inflationary perturbations in the Sec.~\ref{sec3}, when we study the issue of the surviving of PBH cluster. 

The characteristic halo mass is defined as $S_{\rm p}(\Mch,z) = \delta_{\rm c}^2$~\cite{Mo:2002ft}
\begin{equation} \label{M_char}
    \Mch = \frac{9 \, m \fbh }{4 \, \delta_c^2} \left ( \frac{1 + z_{\rm eq}}{1 + z} \right)^{2},
\end{equation}
where $\delta_{\rm c} = 1.69$ is the critical threshold for spherical collapse. It is also implied here that the halo contains both PBHs and dark matter particles, with the mass fraction of PBHs is $\fbh \Mch$. According to the Press-Schechter formalism, the fraction of matter in a halo with a mass greater than $\Mch$ is $\approx 32\%$. The halo mass function is given by~\cite{1974ApJ...187..425P}
\begin{equation} \label{PS_mf}
    M \frac{d n}{d M} = \frac{1}{\sqrt{2 \pi}} \frac{\rho_M}{M} \sqrt{\frac{M}{\Mch}} \exp{\left ( -\frac{M}{2 \Mch} \right)}.
\end{equation}

Forming quite early, such dark matter structures have a high density, so the black holes in them actively interact gravitationally with each other and with dark matter particles. The characteristic timescale of internal evolution for such halos turns out to be much less than the age of the Universe $t_0$. Our previous work~\cite{Stasenko:2023zmf} studied the dynamics of early dark matter structures before core collapse. This paper also takes into account the evolution after core collapse. At this stage of the evolution of a self-gravitating system, the main role is played by binaries, formed in three-body interactions (hereafter we will call them three-body binaries). Some of these binaries will be hard, i.e. their binding energy is greater than the characteristic kinetic energy of the PBHs in cluster. Interactions of such binaries with single PBHs in a cluster increase their binding energy~\cite{1975MNRAS.173..729H}. Thus, binaries become more harden while they pump energy to other PBHs. As a result of such successive scatterings, the binary acquires recoil energy and leaves the core. The heating rate of the cluster per unit mass due to this process is given by~\cite{1987ApJ...319..801L, 1989ApJ...342..814C, 1996PASJ...48..691T}
\begin{equation} \label{E_bheat}
    \dot{E}_{\rm b} = 90 \, G^5 m^3 \rho_{\rm BH}^2\sigma_{\rm BH}^{-7},
\end{equation}
where $\rho_{\rm BH}$ and $\sigma_{\rm BH}$ are the density and one-dimensional velocity dispersion of PBHs. 

\subsection{Simple model for cluster evolution} \label{hom_ev}

To begin with, let us analyze the evolution of the DM halo for the case $\fbh = 1$, i.e. the halo consists only of PBHs. The dynamics of the halo are divided into two parts: before and after the core collapse. Scattering of PBHs on each other occurs most actively in the central part with the highest density, so further considerations refer specifically to the cluster core. The evolution before core collapse occurs under the influence of two-body relaxation is the distant pair scattering of PBHs. As a result, PBHs gradually escape from the core and carry away a certain amount energy. This can be described within the framework of a simple homologous model by the following equation~\cite{1987degc.book.....S}
\begin{equation} \label{dEdt}
    \frac{d E_\c}{dt} = \frac{\zeta E_\c}{M_\c} \frac{d M_\c}{dt},
\end{equation}
here the index $\c$ denotes the core and the numerical value of the constant $\zeta = 0.737$. $M_{\rm c}$ is the mass of the core, the rate of decrease of which is given by
\begin{equation} \label{dMdt}
    \frac{d M_{\rm c}}{dt} = -  \frac{\nu M_{\rm c}}{t_{\rm r}},
\end{equation}
where $t_{\rm r} \sim t_{\rm dyn} N_\c / \ln{N_\c}$ is the relaxation time and $t_{\rm dyn} \sim 1 / \sqrt{G \rho_\c}$ is the core dynamical time. The numerical factor before the relaxation time is somewhat arbitrary and is not important for our analysis here. Applying the virial theorem, we obtain the equation relating the mass and radius of the core
\begin{equation}
    \frac{\dot{r}_{\rm c}}{r_{\rm c}} = \frac{\dot{M}_{\rm c}}{M_{\rm c}} (2 - \zeta).
\end{equation}
After integration we finally get $r_\c(t) \propto M_\c(t)^{2 - \zeta}$, i.e. the shrinking of core occurs with a decrease in mass. The central density evolves as $\rho_\c  \propto M_\c^{-2.79}$. If the evolution proceeds only under the influence of pairwise relaxation, then full core evaporation occurs. Density tends to infinity over time while mass goes to zero, this is a well-known gravothermal catastrophe~\cite{1968MNRAS.138..495L}. 

However, as core density increases, dissipative processes become more and more important. As noted earlier, the core collapse is stopped due to the formation of three-body binaries and their interaction with other PBHs in the cluster. The evolution of the cluster is no longer described only by Eq.~\eqref{dEdt} and it is necessary to take into account the heating of the cluster according to Eq.~\eqref{E_bheat}. The number of PBHs that remain in the core does not depend on the initial conditions. Indeed, at the moment of the core collapse termination, the condition $\dot{E}_\c /M_\c \sim \dot{E}_{\rm b}$ is satisfied. Then we get $N_\c \propto 1 / \ln{N_\c}$ that is approximately a constant up to a logarithmic factor. Note that in the literature it is sometimes (somewhat erroneously) assumed that the cluster completely evaporates, i.e. full core evaporation occurs. However, as follows from the analytical estimation and we will show further as a result of the numerical solution, a finite number of PBHs of the order $N_\c \approx 10 - 30$ remains in the cluster core. Further evolution of the cluster after the core collapse proceeds according to self-similar expansion and is determined by heating due to three-body binaries of Eq.~\eqref{E_bheat} and is described by the following equation 
\begin{equation}
    \frac{G M_c}{r_\c^2} \, \dot{r}_\c = \dot{E}_{\rm b}.
\end{equation}
Since $\sigma_\c \sim \sqrt{G M_\c / r_\c}$, then $\dot{r}_\c \propto r_\c^{-1/2}$, which yields $r_\c \propto t^{2/3 }$ and $\rho_\c \propto t^{-2}$. The presented analysis corresponds to the case of neglecting the contribution of DM particles $\fbh = 1$. The next section considers case $\fbh < 1$ using the kinetic equation. Nevertheless, it can already be noted that the evolution of the cluster is divided into two stages: an increase in the central density until the core collapse; and after collapse dynamics with a decrease in density. 

At the stage of postcollapse evolution, the halo density decreases as $\rho \propto t^{-2}$. The average matter density of the Universe obeys the same dependence. However, at the moment when the cluster enters the expansion stage, its density in the center is higher than at the time of its formation, and therefore significantly higher than the average density of matter in the Universe at this moment. During the structures formation, clusters are accreted into a large halo and the central part of clusters is much denser than halo density. Therefore the cluster is resistant to destruction, this is analyzed in Sec.~\ref{sec3}. 

\subsection{Numerical solution of the Fokker-Planck equation}

To study in more detail the dynamics of early DM halos, especially for the case $\fbh < 1$, we solve the orbit averaged kinetic Fokker-Planck equation in the spherically symmetric case of matter distribution~\cite{1980ApJ...242..765C,1989ApJ...343..725Q, 2017ApJ...848...10V} (see also book~\cite{2013degn.book.....M}). This equation describes the time evolution of the distribution function under the influence of weak pair scattering (two-body relaxation). This approach takes into account the interaction of PBHs with dark matter particles. Namely, PBHs experience dynamical friction against DM particles, which changes the simple core collapse dynamics discussed in Sec.~\ref{hom_ev}. We also assume that the angular momentum distribution is isotropic, i.e. the distribution function depends only on the energy $f(E)$. The method for solving the Fokker-Planck equation is presented in the appendix. We use the kinetic approach, because it is necessary to study the internal dynamics of a huge number of early halos. Direct N-body integration makes it possible to study the dynamics of structures in detail, but it is computationally time consuming. In this context, the Fokker-Planck approach is well suited due to its computational speed while reflecting the main patterns of halo evolution. 

In the previous section it was shown that the internal dynamics of gravitationally interacting bodies ultimately leads to core collapse or, in other words, to a gravothermal catastrophe. In order to take into account after core-collapse dynamics, we modify the Fokker-Planck equation by adding a term responsible for cluster heating associated with three-body binaries, similar to the works~\cite{1987ApJ...319..801L, 1989ApJ...342..814C, 1996PASJ...48..691T}. This is done by the standard procedure of integration along the orbit of Eq.~\eqref{E_bheat}, which is also described in the Appendix. 

We choose the Burkert profile as the initial halo density distribution~\cite{1995ApJ...447L..25B} 
\begin{equation} \label{burkert}
    \rho = \frac{\rho_\c}{\left (1 + r / r_0 \right) \left (1 + r^2 / r_0^2 \right)},
\end{equation}
where the parameter $r_0$ determines the radius of the halo core, for simplicity it is set equal to $r_0 = R_{\rm vir}/5$, that corresponds to $\rho_\c \approx 10^4 \rho_{\rm M}(z)$. We also assume that PBHs and DM particles are equally distributed in space and fraction of PBHs in the halo mass is $\fbh$. The virial radius $R_{\rm vir}$ is determined by the standard expression 
\begin{align}
    R_{\rm vir} &= \left ( \frac{3 M}{4 \pi \Delta_\c \rho_{\rm M}(z)} \right)^{1/3} \nonumber \\
    & \approx 76 \left ( \frac{M}{10^5 \, M_{\odot}} \right )^{1/3} \left ( \frac{1 + z}{20} \right)^{-1} \, \text{pc},
\end{align}
here $\Delta_\c = 18 \pi^2$, $M$ is the halo mass and $\rho_{\rm M}$ is the matter density of the Universe with the modern value is $\rho_{\rm M}(0) \approx 38$~$M_{\odot}$~kpc$^{-3}$.

\begin{figure}
	\begin{center}
\includegraphics[angle=0,width=0.5\textwidth]{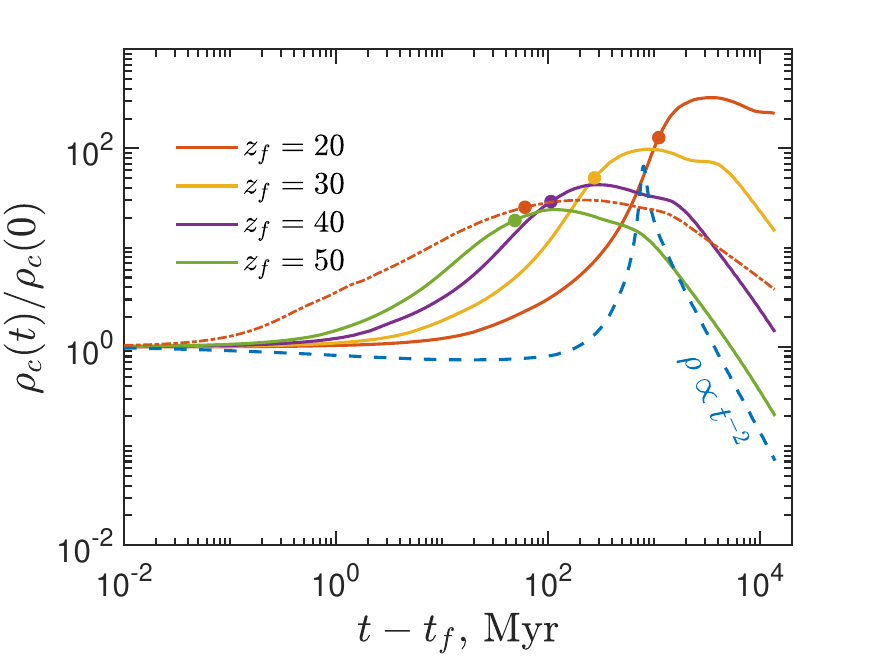}
	\end{center}
\caption{The time dependence of the central density of PBHs in halo with characteristic mass of Eq.~\eqref{M_char}. Solid lines correspond to $\fbh = 0.1$ and different moments of halo formation $z_f$ showed in the legend. The dashed line corresponds to the case $\fbh = 1$ and $z_f = 100$. The red dot-dashed line shows the case $f_{\rm PBH} = 0.01$ and $z_f = 20$. The dots mark are the moments of core collapse from Eq.~\eqref{t_cc}}
\label{gr_rho_c}
\end{figure}

\begin{figure}
	\begin{center}
\includegraphics[angle=0,width=0.5\textwidth]{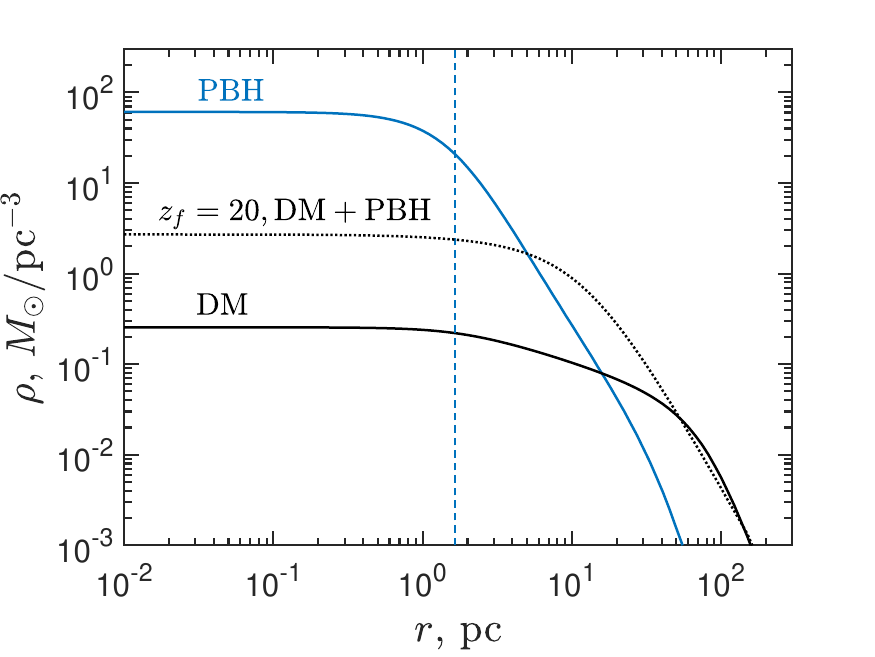}
	\end{center}
\caption{The density profile of PBHs and DM particles at the present moment for a halo of characteristic mass $\Mch$, formed at $z_f = 20$ and $\fbh = 0.1$. The vertical dashed line shows the core radius from Eq.~\eqref{r_core} at $z = 0$. Dotted line is the initial halo density profile at the time of formation}
	\label{gr_rho_zf=20} 
\end{figure}

Figure~\ref{gr_rho_c} shows the evolution of the central density of PBHs (solid lines) for halos of characteristic mass formed at different redshifts for the case $\fbh = 0.1$. Also as an example, the case $\fbh = 1$ with single mass PBHs is shown by dashed line for a halo formed at $z_f = 100$. The dominant presence of dark matter particles alters evolution. If the halo consists only of PBHs, then the central density changes very slowly with time but by the core collapse time grows sharply. For $\fbh < 1$, the initial increase in the PBHs density occurs primarily due to dynamical friction against light DM particles. Also, the core collapse occurs “more softly”. A significant contrast in the PBHs density is formed, which exceeds the initial central density of the halo $\rho = \rho_{\rm PBH} / \fbh$. That is, PBHs over time begin to dominate in density over the dark matter particles in the center of the halo. We call this contrast a PBH cluster. After reaching a sufficiently high density, PBHs will mainly scatter with each other (not DM particles). The further evolution develops along the path of two-body relaxation in accordance with the discussion of the previous section. The dots in Fig.~\ref{gr_rho_c} indicate the core collapse time obtained in~\cite{Stasenko:2023zmf}, where only precollapse dynamics were taken into account
\begin{equation} \label{t_cc}
    t_{\rm cc} = \frac{15.9 \, \sigma^3}{G^2 m \rho_{\c} \ln{\Lambda}} \, \Big (1.3 \, e^{2.1 \fbh} - 1\Big).
\end{equation}
In this work, by core collapse we mean the achievement of the maximum PBH density and, as a consequence, the formation of a cluster. The core collapse time in Eq.~\eqref{t_cc} turns out to be in good agreement with more general our calculations that take into account after core collapse dynamics. Note for the case $f_{\rm PBH} \lesssim 0.1$, the evolution of the halo occurs mainly due to the dynamical friction of the PBHs on the DM particles. The latter causes the core collapse time to scale approximately as $t_{\rm cc} \propto f_{\rm PBH}$ as can be seen from Eq.~\eqref{t_cc} and in Fig.~\ref{gr_rho_c} from the comparison of the red solid line and the dot-dashed line.

Also, for the case $\fbh = 1$, the expansion stage $\rho \propto t^{-2}$ begins immediately after the core collapse. However for the case $\fbh = 0.1$, expansion begins later and the central density remains approximately constant for several core collapse time $t_{\rm cc}$. Next the cluster begins to expand due to the formation of three body  binaries, which act as a source of energy. In our work, we neglect merging of these binaries, because they generally have a lifetime significantly exceeding the age of the Universe~\cite{Franciolini:2022ewd}. Although we note that binaries can acquire a large eccentricity when perturbed by a third body due to the Kozai-Lidov mechanism \cite{LIDOV1962719, 1962AJ.....67..591K}. Particularly in globular clusters in hierarchical triple systems, the inner binary can merge in Hubble time due to this process. \cite{Wen:2002km}. However, we assume that this effect is very rare for our consideration and is beyond the scope of this work. The asymptotic behavior of the cluster expansion $\rho \propto t^{-2}$ does not take place for $\fbh < 1$, although the dependence is similar, that is due to the abundant presence of dark matter particles, which also interact with PBHs. In addition, for halos formed at $z_f \gtrsim 30$, the expansion of the cluster leads to the fact that its density at $z = 0$ turns out to be less than at the moment of formation. 

In Fig.~\ref{gr_rho_zf=20} we show the density profile of PBHs and DM particles at the present moment $z = 0$. The dotted line is the halo density profile at the moment of formation $z_f$ = 20, and the fraction of PBHs is $\fbh$. The vertical dashed line shows the core radius (also known as King radius), which is given by~\cite{1966AJ.....71...64K, 2008gady.book.....B}
\begin{equation} \label{r_core}
    r_{\c} = \sqrt{\frac{3 \, v^2(0)}{4 \pi G \rho(0)}},
\end{equation}
here $v(0)$ is the three-dimensional velocity dispersion at the center of the halo, which in a multicomponent environment can be defined as~\cite{1989ApJ...343..725Q}
\begin{equation}
    v^2(0) = \frac{1}{\rho(0)} \sum \rho_i v^2_i(0),
\end{equation}
where $v_i$ is given by
\begin{equation}
    v^2_i = \frac{4 \pi m_i}{\rho_{i}} \int_0^{\phi^{-1}(E)} dE \, f_{i}(E) \Big [2 (E - \phi(r)) \Big]^{3/2},
\end{equation}
in our case, the index $i$ refers to either DM particles or PBHs. With this definition, at point $r_\c$ the density turns out to be three times less than the central one. We also note that three-dimensional and one-dimensional velocity dispersion are related by the expression $v^2 = 3 \sigma^2$.

Our calculations show that about $N_\c \simeq 20$ PBHs remain in the core. From Fig.~\ref{gr_rho_zf=20} it is clear that the PBHs density at $z = 0$ exceeds the halo density at the initial time at $z_f = 20$. In addition, the density of DM particles is approximately two orders of magnitude lower than the PBHs in the center of the halo, which is due to mass segregation. It can be seen that the region of PBHs dominance is order of $\sim 10$~pc. Thus, the internal evolution of dark halos leads to the formation of dense PBH clusters. 

To conclude this section, we will briefly discuss the presence of baryonic matter in the halo. The infalling gas is thermalized to a virial temperature
\begin{equation}
    T_{\rm vir} = \frac{\mu m_p}{2 k_B} \frac{G M}{R_{\rm vir}} \approx 300 \left ( \frac{M}{10^{5} \, M_{\odot}} \right)^{2/3} \left ( \frac{1 + z}{20} \right) \, \rm{K},
\end{equation}
here $\mu$ is the mean molecular weight and $m_p$ is the proton mass. For baryonic gas dynamics to differ from dark matter, the baryonic component should cool. Atomic cooling is effective only at temperatures $T_{\rm vir} > 10^4$~K. Therefore, the behavior of baryons is no different from dark matter particles. However, the accretion of matter onto PBHs will lead to heating and ionization of the matter. Inverse compton scattering of CMB photons by electrons will effectively cool the gas. In order to understand the influence of baryonic effects, it is necessary to know the efficiency of accretion. Because of the complexity of this analysis, we assume that baryons are indistinguishable from dark matter particles. In other words, we neglect the influence of baryons on the evolution of the halo. 

\section{The surviving of clusters} \label{sec3}

In the process of hierarchical structure formation, small halos will be absorbed into larger halos and become subhalos. To simplify the terminology, we will assume that the cluster and subhalo are identical. The large halo in which the cluster is located will be called the host halo. The characteristic timescales for the structure formation correspond to the formation time of the halo $t_f$ (i.e., Hubble time), because the halo mass in Eq.~\eqref{M_char} changes as $\Mch \propto (t/t_f)^{4/3}$. Hence, within a time of the order of $t_f$, the halo already begins to merge and form a larger halo. 

The probability that a halo with mass $M_1$, formed at time $t_1$, will be in a halo with mass $M_2$ by time $t < t_2$ is given according to the Extended Press-Schehter formalism by the expression~\cite{1993MNRAS.262..627L}
\begin{align} \label{P_abs2}
    &P(S_2,t < t_2 | S_1,t_1) = \frac{1}{2} \Big [1 - \rm{erf}(A) \Big] + \nonumber \\
    & \, \, \frac{\delta_1 - 2\delta_2}{2 \delta_1} \exp{\left (\frac{2 \delta_2 (\delta_1 - \delta_2)}{S1} \right)} \Big [1 - \rm{erf}(B) \Big],
\end{align}
here
\begin{align}
    A &= \frac{S_1 \delta_2 - S_2 \delta_1}{\sqrt{2S_1 S_2(S_1 - S_2)}}, \nonumber \\
    B &= \frac{S_2 (\delta_1 - 2 \delta_2) + S_1 \delta_2}{\sqrt{2S_1 S_2(S_1 - S_2)}},
\end{align}
where $S = S_{\rm p} + S_{\rm ad}$ is the modern dispersion of fluctuations. In addition to the Poisson noise of the PBHs, we also took into account adiabatic inflation perturbations, which are well approximated by the following formula~\cite{2011itec.book.....G}
\begin{equation} \label{sigma_ad}
    S_{\rm ad}(M,z) \approx \frac{0.043}{(1+z)^2} \ln^{2.5} \left ( \frac{2.3 \times 10^{15} \, M_{\odot}}{M} \right),
\end{equation}
$S_1 = S(M_1)$, $S_2 = S(M_2)$ and $\delta = \delta_\c (1+z)$ is the threshold for collapse as a function of time. We neglect the effects of the cosmological constant $\Lambda$ in the linear growth of fluctuations, because we consider redshifts where its influence is negligible. 

\begin{figure}
\centering
\includegraphics[width=0.5\textwidth]{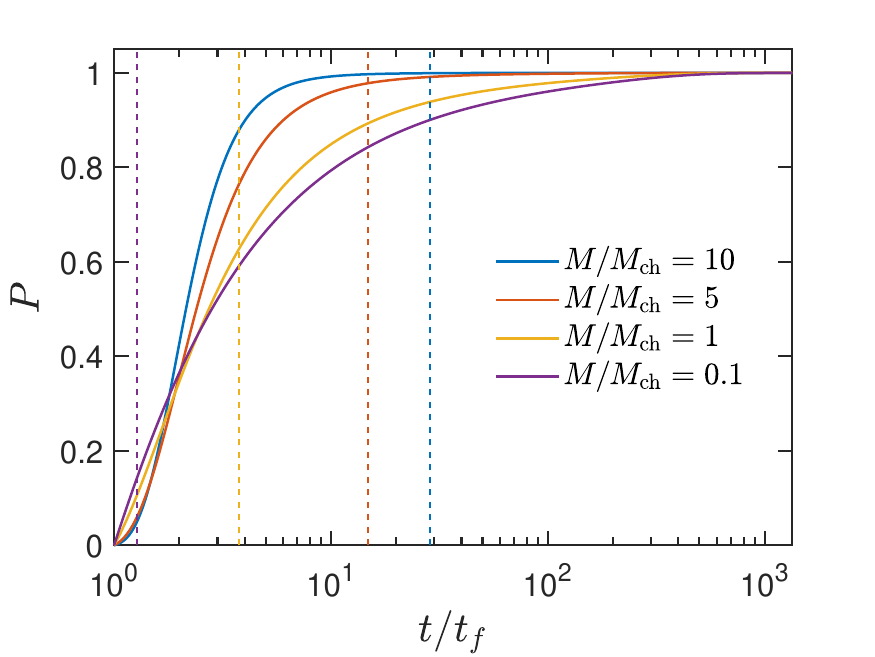}
\caption{The cumulative probability of absorption of a halo with mass $M_1$, marked in the legend, into a halo with mass $M_2 = 2M_1$. The graph is shown for the moment of formation halos $z_f = 30$ and PBHs fraction $\fbh = 0.1$. The dashed vertical lines show the core collapse time (the formation of PBHs cluster), their color matches the color of the solid lines.}
\label{P_ab}
\end{figure}

Following Ref.~\cite{1993MNRAS.262..627L}, we define the halo survival time as the time during which the halo mass doubles. Figure~\ref{P_ab} shows the cumulative probability that a halo with mass $M_1$ will be in a halo with mass $2 M_1$ as a function of time. Here we set $\fbh = 0.1$ and the moment of halo formation $z_f = 30$. In the Eq.\ \eqref{P_abs2} we assume that $t_1 = t_f$ $t_2 = t/t_1$. We note that the form of the function $P(t/t_f)$ weakly depends on the specific moment of formation. 

The vertical lines in Fig.~\ref{P_ab} show the time of core collapse for halos of different masses in accordance with the color in the legend. Let us remind that the time of core collapse corresponds to the time of formation of a PBH cluster. It can be seen that the core collapse in a halo of small masses predominantly occurs before their mass doubles. Thus, a PBHs cluster is more likely to form in low-mass halos. Also, for halos that form earlier, the time of the core collapse turns out to be shorter as compared to the time of formation 
\begin{equation}
    \frac{t_{\rm cc}}{t_{f}} \propto \frac{\sigma^3}{\sqrt{\rho}} \propto (1 + z)^{-2}.
\end{equation}
Therefore, the earlier a halo is formed, the more efficiently PBH clusters are formed in them. Note also that the case $\fbh \approx 1$ is disfavoured, because collapse occurs over a large number of times $t_f$, as follows from Eq.~\eqref{t_cc}. In particular, from Fig.~\ref{gr_rho_c} it follows that for $\fbh = 1$ and $z_f = 100$ the core collapse of the characteristic mass halo occurs in $t_{\rm cc} \approx 40 \, t_f$. In addition, the collapse is sharp, i.e.\ it occurs at the end of pre-collapse evolution. The dense PBH cluster is not formed in such structures, because absorption into a large halo occurs much earlier than core collapse. On the other hand for PBH fraction $\fbh = 0.1$, the halo of characteristic mass, formed at $z_f = 30$, undergoes the core collapse in $t_{\rm cc} \approx 4 \, t_f$. 

We are also interested in merging of PBHs in clusters, so it is necessary to understand how many clusters survive by the modern era. After absorption into a large halo, under the influence of dynamical friction, a subhalo with mass $M$ settles in the host halo in time
\begin{equation} \label{t_df_cl}
    t_{\rm df} = \frac{400\, \text{Gyr}}{\ln{\Lambda}} \left ( \frac{R}{1 \, \text{kpc}} \right)^{2} \left ( \frac{\sigma_{\rm h}}{10 \, \text{km} \, \text{s}^{-1}} \right ) \left ( \frac{10^4 \, M_{\odot}}{M} \right ),
\end{equation}
where it is assumed that the host density profile has the form of a singular isothermal sphere $\rho = \sigma_{\rm h}^2 / 2\pi G r^2$ with velocity dispersion $\sigma_{\rm h}$ and $R$ is the radius from which dynamical friction sink begins. Equation~\eqref{t_df_cl} also implies that the cluster sinks in a circular orbit. For orbits with low angular momentum, the dynamical friction time will be less~\cite{vandenBosch:1998dc}. In modern halos, dynamical friction is ineffective for clusters with low mass, but in earlier epochs this was not the case. Further, at the end of this section, we will show that if a cluster is absorbed by a host halo at redshifts $z > 4$, then it settles in time less than the age of the Universe. 

During the process of settling, the subhalo will experience gradual tidal striping under the influence of tidal forces from the host halo potential. The mass is located at a distance greater than the tidal cutoff radius $r_{\rm t}$ from the center of the cluster is captured in the host~\cite{vandenBosch:2017ynq} 
\begin{equation}
    r_{\rm t} = R \left ( \frac{M(r_{\rm t})}{M_{\rm h}(R) \Big [3 - \frac{d \ln M_{\rm h}}{d \ln r}|_{r = R} \Big]} \right)^{1/3},
\end{equation}
where $R$ is the radius of the circular orbit in which the cluster rotates. In fact, $r_{\rm t}$ refers to the radius of the subhalo. For an host halo isothermal sphere, the tidal radius will be
\begin{equation}
    r_{\rm t} = R \left ( \frac{M}{2 M_{\rm h}(R)} \right)^{1/3}.
\end{equation}
In order to the cluster to fully experience tidal striping, it is necessary that the central density of the cluster be comparable to the host halo density $\rho_{\rm h}(R) \approx \rho(0)$. However, as we discussed earlier, the cluster is significantly denser than the host, as a consequence of its internal evolution. Therefore, tidal striping is ineffective for the complete destruction of clusters, despite the fact that the cluster can settle into the center of the halo in a short time. 

Another process that can lead to the destruction of clusters is their tidal interaction with each other in the host halo. We will assume that, under the influence of dynamical friction, the clusters settled in the center of host halo, where they then interact with each other. During long-range tidal interactions of clusters, the rate of change in their internal energy is given by~\cite{2008gady.book.....B}
\begin{equation} \label{E_tid_h}
    \dot{E}_{\rm tid} \approx \frac{14}{3} \sqrt{2 \pi} \frac{G^2 M^3 n_{\rm cl} r^2}{\sigma_{\rm h}} \int_r^{\infty} \frac{db}{b^3},
\end{equation}
here $n_{\rm cl}$ is the number density of clusters in the halo, which, to maximize the effect, we define as $n_{\rm cl} = \rho_{\rm h}/ M$. We estimate the time required to destroy a cluster (tidal disruption time) as $t_{\rm td} \sim E_{\rm cl} / \dot{E}_{\rm tid}$
\begin{equation}
    t_{\rm td} \sim \frac{0.2}{\sqrt{G \rho_{\rm h}}} \left ( \frac{\sigma_{\rm h}}{\sigma} \right ) \sqrt{\frac{\rho}{\rho_{\rm h}}}.
\end{equation}
It can be seen that the destruction of clusters lasts several central dynamic times of the host halo. However, this is a conservative estimation of the tidal disruption time and in a realistic case it can be much longer. Thus, if a cluster turns out to be in a halo, where its dynamical friction time is less than the age of the Universe, then we believe that it is being destroyed by the present moment. 

At low redshifts, structures are formed from inflationary perturbations, because Poisson noise turns out to be weak. Let us use Eq.~\eqref{t_df_cl} to estimate the dynamical friction time of a subhalo of mass $M$ in a halo of characteristic mass $S_{ad}(\Mch,z) = \delta^2_c$ 
\begin{equation} \label{tdf2}
    t_{\rm df} = 15 \, \frac{e^{\xi(4)}}{e^{\xi(z)}} \left ( \frac{1+z}{5} \right)^{-3/2} \left ( \frac{10^4 \, M_{\odot}}{M} \right ) \, \rm{Gyr}
\end{equation}
here for simplicity we have introduced the function
\begin{equation}
    \xi(z) = \Big (5 \delta_\c (1+z) \Big)^{0.8},
\end{equation}
we put $\ln{\Lambda} = 10$ and the distance from which the sink begins $R = R_{\rm vir}$. If the cluster is absorbed at low redshifts, then the dynamical friction time will exceed the age of the Universe. In particular, the dynamical friction becomes ineffective at $z \lesssim 4$ for clusters with mass $M \sim 10^4 \, M_{\odot}$. If the cluster survives to the epoch where dynamical friction becomes ineffective, then we will assume that the cluster is not further destroyed. 

\begin{figure}
	\begin{center}
\includegraphics[angle=0,width=0.5\textwidth]{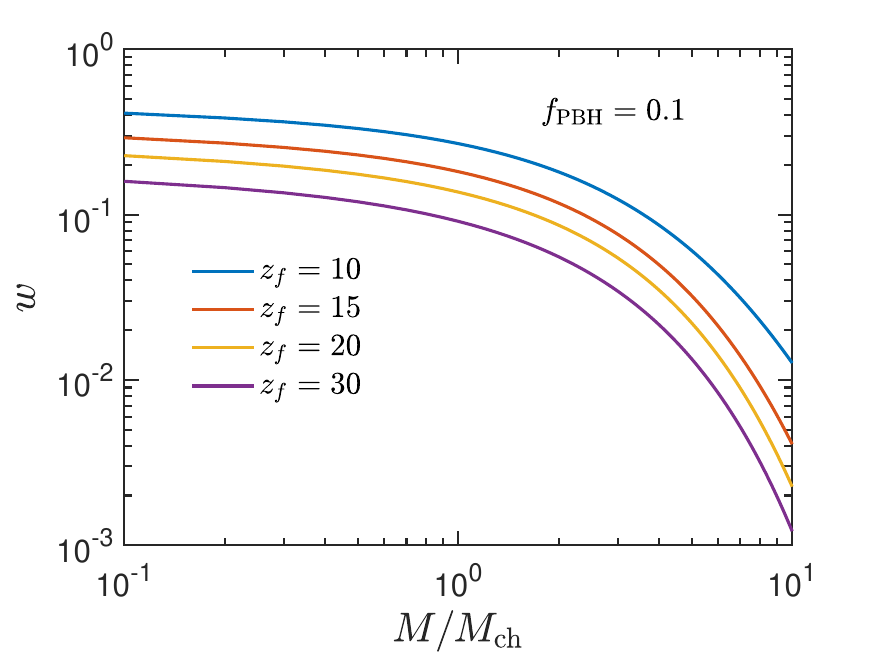}
	\end{center}
\caption{The fraction of surviving clusters by the modern epoch $z = 0$ depending on their mass. Different lines correspond to different moments of formation $z_f$ showed on the legend.}
	\label{surv_frac} 
\end{figure}

Let us determine the fraction of halos $w$ that survive the process of structure formation to the modern era. We will proceed as follows: for a cluster with mass $M$, formed at $z_f$, we find the moment $z_{ s}$ from Eq.\ \eqref{tdf2}, at which the time of dynamical friction is comparable with the age of the Universe $t_0$. Then, using Eq.\ \eqref{P_abs2}, we find the probability $P_{s}$ that the cluster mass has doubled by time $z_{ s}$, the required fraction will be $w = 1 - P_{s}$. If the cluster survived until the moment $z_{s}$, then further absorption into large halos will not lead to its destruction. Fig.~\ref{surv_frac} shows the fraction of surviving clusters $w$ for different moments of their formation $z_f$. 

Let us estimate the fraction of matter that will be in a halo with a mass greater than $M$ by the present moment
\begin{equation} \label{cl_frac1}
    F = \frac{1}{\rho_{\rm M}} \int_{M}^{\infty} M' dN(M',z_f) w(M',z_f),
\end{equation}
where $dN$ is given by Eq.\ \eqref{PS_mf}. As can be seen from Fig.~\ref{surv_frac} the function $w$ weakly changes up to halo mass $\lesssim \Mch$ and sharply decreases with $M > \Mch$. Also, integral in Eq.~\eqref{cl_frac1} weakly depends on lower limit, so we change the lower and upper limits by $0$ and $\Mch$ respectively. As expected, the fraction of matter that will remain in the early structures is $F \sim w$, where the maximum value of the function $w$ from Fig.~\ref{surv_frac} is taken for estimation. However, most of the subhalo mass is stripped out into the host halo, but the central part of the subhalo, containing a dense PBH cluster, probably survives~\cite{Delos:2019lik}. Thus, about $10 - 40\%$ of emerging clusters survive during the structure formation and can be widespread in the modern Universe.

Let us make an important note: the doubling of the halo mass mainly occurs due to the absorption of halos of lower masses. Therefore, this does not necessarily mean the destruction of the PBH cluster. It is likely that the influence of this process on the internal evolution of the cluster is negligible. However, for a conservative estimate, we consider a doubling of the halo mass to be formally fact of the cluster destruction.

\section{Mergers of PBHs} \label{sec4}

In this section, we separately consider mergers of PBH binaries formed both in the early Universe before $z_\eq$ and deep inside the matter-dominated era as a result of close pair scatterings in dark halos. As was shown in the previous section, the evolution of early DM halos leads to the formation of a dense PBH clusters, in which the probability of PBHs mergers becomes enhanced in comparison with the nonevolutionary case. 

\subsection{Early binaries}

A pair of PBHs, due to their initial Poisson distribution, forms a binary system at the RD stage. This pair gradually shrinks due to the emission of GW waves and the rate of loss of energy and angular momentum is given by~\cite{PhysRev.136.B1224}
\begin{align}
    \dot{E}_{\rm gw} &= - \frac{64 \, G^4 m^5}{5 \, c^5 a^5 (1 - e^2)^{7/2}} \left (1 + \frac{73}{24}e^2 + \frac{37}{96} e^4 \right), \label{E_gw_dot} \\
    \dot{L}_{\rm gw} &= - \frac{32 \sqrt{2} G^{7/2} m^{9/2}}{5 c^5 a^{7/2} (1 - e^2)^2} \left ( 1 + \frac{7}{8} e^2 \right) \label{L_gw_dot},
\end{align}
where $a$ is the semimajor axis and $e$ is the eccentricity. The lifetime of a binary due to the emission of gravitational waves is given by 
\begin{equation} \label{t_gw}
    t_{\rm mer}=\frac{3 \, c^5 a^4 j^7}{170 \, G^3m^3},
\end{equation}
where we introduced dimensionless angular momentum $j = \sqrt{1 - e^2}$. Note that binaries formed in the early Universe are highly eccentric $j \ll 1$~\cite{2017PhRvD..96l3523A}. The PBH merger rate as a function of cosmic time is given by~\cite{Raidal:2018bbj}
\begin{align} \label{mr0}
    \mathcal{R}_0 &= \frac{3.1 \times 10^6}{\text{Gpc}^{3} \, \text{yr}} \,  \fbh^{53/37} \, \left ( \frac{t}{t_0} \right)^{-34/37} \left ( \frac{m}{ M_{\odot}} \right)^{-32/37} \nonumber \\
    & \times 0.24 \left ( 1 + \frac{2.3 \, S_{\rm eq}}{\fbh^2} \right)^{-21/74}
\end{align}
where $t_0$ is the age of the Universe. The last term reflects the influence of adiabatic perturbations on the eccentricity of early binaries~\cite{Eroshenko:2016hmn, 2017PhRvD..96l3523A} and $\sqrt{S_{\rm eq}} \approx 0.005$ is the variance of matter perturbation at matter-radiation equality. 

The merger rate in Eq.~\eqref{mr0} does not take into account the perturbation of the binary parameters due to interactions with other PBHs in early halos. The probability that the halo contains $N$ PBHs is given by~\cite{Inman:2019wvr}
\begin{equation}
    p_N \propto \frac{1}{\sqrt{N}} e^{-N/N_*(z)},
\end{equation}
where
\begin{equation} \label{Nchar}
    N_* = \left ( \ln(1 + \delta_*) - \frac{\delta_*}{1 + \delta_*} \right)^{-1}
\end{equation}
here $\delta_* = \delta_c / ( D(z) \fbh)$ and $D(z)$ is given by Eq.~\eqref{Dz}. At redshifts $z \ll z_{\rm eq}$ deep in the matter domination stage $\delta_* \ll 1$, then the Taylor expansion in Eq.~\eqref{Nchar} \ gives $N_* \approx 2 / \delta_*^ 2$ which will be $2 \Mch \fbh / m$ as it should be for the Press-Schechter mass function in Eq.~\eqref{PS_mf} in the discrete limit.

According to the analysis of Ref.~\cite{Vaskonen:2019jpv}, the merger rate is modified as $\mathcal{R} = \mathcal{R}_0 P_{\rm np}$, where $P_{\np}$ is an suppression factor showing the fraction of unperturbed binaries. We assume that perturbed binaries do not merge by the present moment. The mechanism for perturbing binaries is that on the timescale of the halo core collapse, the angular momentum of the binary $j$ increases due to scattering with other PBHs. Since early binaries have $j \ll 1$, this leads to a significant increase in the lifetime according to Eq.~\eqref{t_gw}. However, it should be noted that in some rare cases a single PBH can perturb a binary one so that the eccentricity, on the contrary, increases. However, binaries experience several scatterings with single PBHs, and one can expect that on average the angular momentum of binaries will still increase. We defer this analysis to future work. Thus, if the binary is located in a halo that experiences core collapse at redshift $z$, then it will not contribute to the merger rate. Therefore, the fraction of unperturbed binaries will be 
\begin{align} \label{Pnp}
    P_{\np}(z) = &1 - \sum_{N = 3}^{N_{\rm crit}(z)} p_N(z_f) \nonumber \\
    & - \sum_{N' > N_{\rm crit}(z)} \left ( \sum_{N = 3}^{N_{\rm crit}(z)} \widetilde{p}_N(z_f) \right) p_{N'}(z_f)
\end{align} 
where $N_{\rm crit}(z)$ is the critical amount of the PBHs in the halo, which formed at redshift $z_f$ and undergoes the core collapse by $z$. Note that halos containing PBHs $N < N_{\rm crit}$ experienced core collapse at redshifts greater than $z$. We determine the core collapse time according to Eq.\ \eqref{t_cc}. In Eq.\ \eqref{Pnp} the second term is the probability that the binary is in a halo with $N < N_{\rm crit}$ PBHs. The third term corresponds to the probability that, in a halo with a number of PBHs $N > N_{\rm crit}$, the binary will be inside a subhalo of lower mass. Since $P_\np$ depends on $z_f$, we numerically find the moment of halo formation at which $P_\np$ is minimal. Probabilities $p_N$ and $\widetilde{p}_N$ are normalized as follows 
\begin{equation}
    \sum_{N \geq 2} p_N = 1, \, \, \, \, \sum_{N=2}^{N'} \widetilde{p}_N = 1.
\end{equation}

\begin{figure}
	\begin{center}
\includegraphics[angle=0,width=0.5\textwidth]{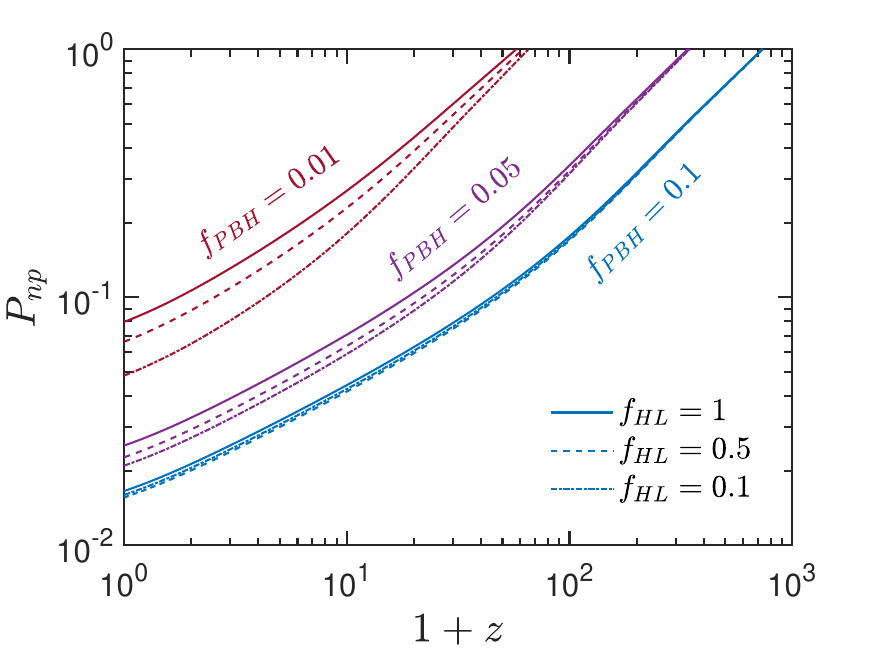}
	\end{center}
\caption{The fraction of unperturbed binaries (suppression factor) as a function of redshift for different fractions of PBHs in the DM $\fbh$ (color lines) and different fractions of DM $f_{\rm HL}$ in the halo composition is given by Eq.~\eqref{M_fhl} in accordance with the marking in the legend.}
	\label{Sz_fpbh} 
\end{figure}

Figure~\ref{Sz_fpbh} shows the dependence of the suppression factor $P_\np$ on the redshift for different $f_{\rm PBH}$ and $f_{\rm HL}$. Different curves in the graph correspond to different fraction of DM particles clustering in the halo, where we parameterized this as in the work~\cite{Inman:2019wvr}
\begin{equation}\label{M_fhl}
    M = m N_{\rm PBH} \left [1 + f_{\rm HL} \frac{1 - \fbh}{\fbh} \right].
\end{equation}
Here $M$ is the mass of the halo. Note $f_{\rm HL} = 1$ corresponds to the fact that the mass of DM particles in the halo is $M(1 - \fbh)$ (the case when there is no biasing in DM and PBH distributions). Another limit $f_{\rm HL} = 0$ means that the halo consists only of PBHs. In Ref.~\cite{Inman:2019wvr} using N-body, it was also shown that $f_{\rm HL}$ increases with time and, in particular, by the time $z = 100$ it is $f_{\rm HL} = 0.4 $ for $\fbh = 0.1$. But since the suppressive factor weakly depends on $f_{\rm HL}$, then in what follows we assume $f_{\rm HL}=1$. 

Let us now qualitatively discuss the perturbation of binaries in the context of the hierarchical structure formation. From the point of view of suppressing mergers of early binaries, the case $\fbh < 1$ is preferable, because the core collapse of such halos occurs within several times $t_f$ characteristic of the structure formation as was shown in Sec.~\ref{sec3} (see Fig.~\ref{P_ab}). The binaries in such halos will have time to become perturbed before the cluster is absorbed by a large halo. On the contrary, in the case $f_{\rm PBH} = 1$ with monochromatic mass spectrum of PBHs, the time scale of two-body relaxation significantly exceeds the characteristic time for the structure formation. Therefore, the internal structure of the cluster (and binaries) will not change from the moment of its formation until absorption into a large halo.

\subsection{Late binaries}

As was shown earlier, the evolution of early DM halos leads to the formation of dense PBH clusters. In these clusters, two PBHs can approach each other quite closely and form a pair due to the emission of gravitational waves. In this section, we derive the cross section for binary formation and estimate its parameters. 

The amount of the gravitational wave energy emitted in a hyperbolic orbit can be obtained using Eq.~\eqref{E_gw_dot} for elliptic orbit. The main amount of energy is emitted when passing the pericenter $r_\p = a (1 - e)$. Then, during one orbital period $T = \pi a^{3/2} \sqrt{2 / G m}$ in a highly elongated elliptical orbit $e \approx 1$ energy $\delta E \approx \dot{E}_{\rm gw} T$ will be emitted 
\begin{equation} \label{del_E_orb}
    \delta E = \frac{85 \pi \, G^{7/2} m^{9/2}}{12 \, c^5 r_\p^{7/2}}.
\end{equation}
The value $r_\p$ is related to the impact parameter $b$ by the following expression 
\begin{equation} \label{b_1}
    b^2 = r_\p^2 \left (1 + \frac{4 \, G m}{r_\p v_{\rm rel}^2} \right ) \approx \frac{4\, r_\p G m}{v_{\rm rel}^2},
\end{equation}
$v_{\rm rel}$ is the relative velocity between PBHs at infinity. For the formation of a binary, it is necessary that the amount of energy $\delta E = \alpha \mu v_{\rm rel}^2 /2$ be emitted, where $\alpha \geq 1$ and $\mu = m/2$ is the reduced mass. Then from Eqs.~\eqref{b_1} and~\eqref{del_E_orb} we get that the impact parameter is
\begin{align} \label{b_2}
    b &= \frac{2 G m}{v_{\rm rel}} \left ( \frac{85 \pi}{3 \alpha c^5 v_{\rm rel}^2} \right)^{1/7} \nonumber \\
    &\approx 8.9  \left ( \frac{v_{\rm rel}}{ \text{km} \, \text{s}^{-1}} \right )^{-9/7} \left (  \frac{10}{\alpha} \right)^{1/7} \, \text{au},
\end{align}
where PBH mass is $m = 30 \Msun$. In order to obtain the cross section for the formation of a binary, it is sufficient to set $\alpha = 1$ in this case $b_{\rm max} \approx 12.4 \,$~au
\begin{equation} \label{cs_bbh}
    \Sigma = \pi R^2_{\rm s} \left( \frac{85 \pi}{3} \right )^{2/7} \left ( \frac{v_{\rm rel}}{c} \right)^{-18/7},
\end{equation}
here $R_{\rm s} = 2 \, G m / c^2$ is the Schwarzschild radius of the black hole, similar expression for the cross section was obtained in Refs.~\cite{1989ApJ...343..725Q,Mouri:2002mc}.

Let us estimate the parameters and lifetime of the binaries formed through this channel. It follows from Eq.~\eqref{b_2} that the impact parameter very weakly depends on $\alpha$ and $b \approx b_{\rm max}/\sqrt{2}$ for $\alpha = 10$. Therefore, the initial energy of two PBHs $\mu v_{\rm rel}^2/2$ is much less than the energy emitted in the form of gravitational waves $\delta E$. Thus, most of the binaries will have binding energy $G m^2 / 2 a \approx \delta E$. Hence the semimajor axis in turn is determined by the expression
\begin{align}
    a &= \frac{3\, c^5 (b v_{\rm rel})^{7}}{5440 \pi \, G^6 m^6} \nonumber \\  
    &= 5.8 \times 10^3 \left ( \frac{b}{9 \, \text{au}} \right)^{7} \left ( \frac{v_{\rm rel}}{\text{km} \, \text{s}^{-1}} \right)^{7} 
    \, \text{au}.
\end{align}
The average distance between PBHs in a cluster is $d \sim n_{PBH}^{-1/3} \sim 0.8$~pc for PBHs density $\rho_\c \approx 100 \, \Msun$~pc$^{-3}$ (see Fig.~\ref{gr_rho_zf=20}). Thus, binary separation will be approximately two orders of magnitude smaller than the average distance between PBHs in the cluster core. 

The angular momentum of the forming binary reads
\begin{equation}
    L = \mu b v_{rel} + \delta L,
\end{equation}
where the first term is the initial angular momentum $L_0 = \mu b v_{\rm rel}$ and, as in the case of energy, from Eq. \eqref{L_gw_dot} $\delta L \approx \dot{L}_{\rm gw} T$ 
\begin{equation}
    \delta L = \frac{96 \pi G^5 m^6}{c^5 (bv_{\rm rel})^4}.
\end{equation}
It is clear that $\delta L \ll L_0$, because $\sqrt{G m /b} \ll c$. Hence the eccentricity of the formed binary is 
\begin{equation}
    e = \sqrt{1 - \frac{4 \, L_0^2 \delta E}{G^2 m^5}}.
\end{equation}
The dimensionless angular momentum $j = \sqrt{1 - e^2}$ in turn reads 
\begin{align}
    j &= \sqrt{\frac{2720 \pi}{3}} \left ( \frac{Gm}{c b v_{\rm rel}} \right)^{5/2} \nonumber \\
    &= 5.1 \times 10^{-4} \left ( \frac{b}{9 \, \text{au}} \right)^{-5/2} \left ( \frac{v_{\rm rel}}{ \text{km} \, \text{s}^{-1}} \right)^{-5/2}.
\end{align} 
As expected, the late binaries turn out to be incredibly eccentric. The merger time of such binaries from Eq.~\eqref{t_gw} is
\begin{equation}
    t_{\rm mer} \approx 10^5 \left( \frac{b}{9 \, \text{au}} \right)^{21/2} \left ( \frac{v_{\rm rel}}{ \text{km} \, \text{s}^{-1}} \right)^{21/2} \, \text{yr}.
\end{equation}
Late binaries, like early ones, can be destroyed due to strong scattering with other PBHs in the cluster. The characteristic time for this is given by~\cite{Stasenko:2023zmf}
\begin{equation}
    t_{\rm s} \approx 10^7 \left ( \frac{10^4\text{au}}{a} \right ) \left (\frac{100 \, M_{\odot} \, \text{pc}^{-3}}{\rho_{\rm PBH}} \right) \left ( \frac{v_{\rm rel}}{ \text{km} \, \text{s}^{-1}} \right) \, \text{yr}
\end{equation}
Thus, the forming binaries merge before possible destruction due to strong scattering. They also have a short lifetime, so they can be considered to merge instantly after formation. 

\subsection{The PBH merger rate} 

The merger rate of PBHs per halo is
\begin{equation} \label{cl_mr}
    \Gamma_{\rm h}  = \frac{2 \, \pi}{m^2} \int d r \, r^2 \rho_{\rm PBH}^2 \langle \Sigma v_{rel} \rangle,
\end{equation}
where the angle brackets mean the average over the relative velocities of PBHs, which we calculate according to~\cite{1989ApJ...342..814C, Stasenko:2021vmm} and $\Sigma$ is the cross section for the binary formation, which is given by Eq.~\eqref{cs_bbh}. Let us recall, that we assume that after formation, late binaries are instantly merged. The internal dynamics of the halo leads to the fact that the rate of PBHs mergers also evolves over time. The PBH merger rate is obtained by convolving expression in Eq.~\eqref{cl_mr} with the halo mass function in Eq.~\eqref{PS_mf}. It is also necessary to take into account the destruction of the halo during the hierarchical formation of structures. To obtain the correct merger rate, it is necessary to take into account PBH mergers that occur in halos formed at different redshifts, and also to know how many halos are destroyed. For simplicity, we will assume that all halos are formed at some specific $z_f$ and the effects of halo mergers and destructions are described by the function $w$ introduced in Section~\ref{sec3}. The function $w$ shows the fraction of surviving halos during the structure formation. Thus, the merger rate of late binaries is 
\begin{equation} \label{mr_l1}
    \mathcal{R}_{\rm l}(z) = \int w(M,z_f) \frac{dn}{dM}(z_f) \Gamma_{\rm h}(z) \, dM,
\end{equation}
we emphasize that this expression depends on $z_f$. Thus, we assume that after formation at some redshift $z_f$ the mass function is “frozen” and the merger rate changes over time only due to the internal evolution of the halos and their survival. We do not take into account newly formed halos, and the contribution to the merger rate of destroyed halos is zero. Let us remind that according to the Section~\ref{sec3}, we consider a halo undestroyed if it has survived to the moment when dynamical friction is ineffective in large halos. 

\begin{figure}
	\begin{center}
\includegraphics[angle=0,width=0.5\textwidth]{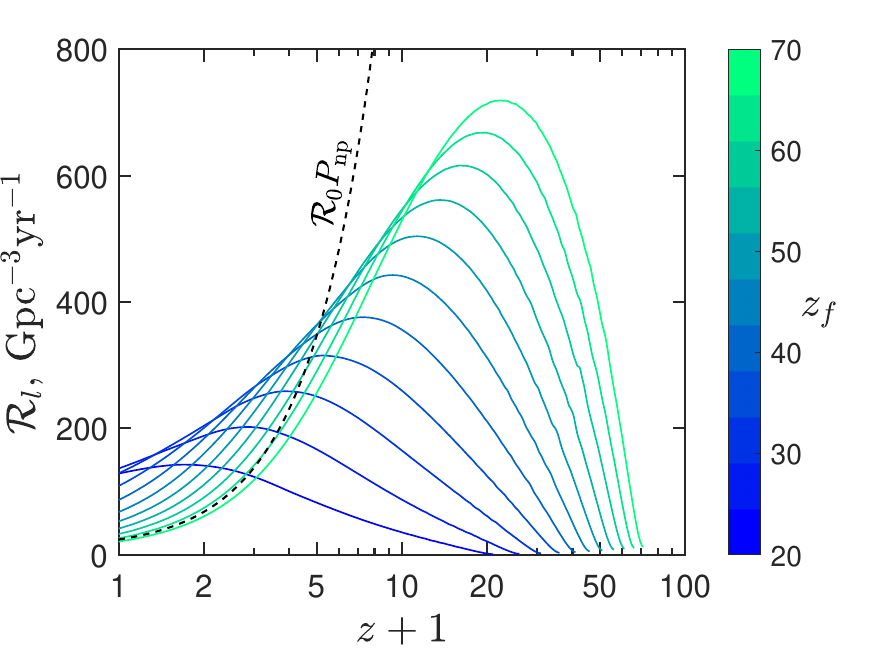}
	\end{center}
\caption{The merger rate of late binaries for the case  $\fbh = 0.1$ and $w = 1$ in Eq.\ \eqref{mr_l1}. The curves from right to left correspond to a decrease of the moment halo formation starting from $z_f = 70$ with a step of $\Delta z = 5$, also displayed on the colorbar. The dashed black curve shows the merger rate of unperturbed early binaries}
	\label{mr_late_diff_z_f} 
\end{figure}

\begin{figure}
	\begin{center}
\includegraphics[angle=0,width=0.5\textwidth]{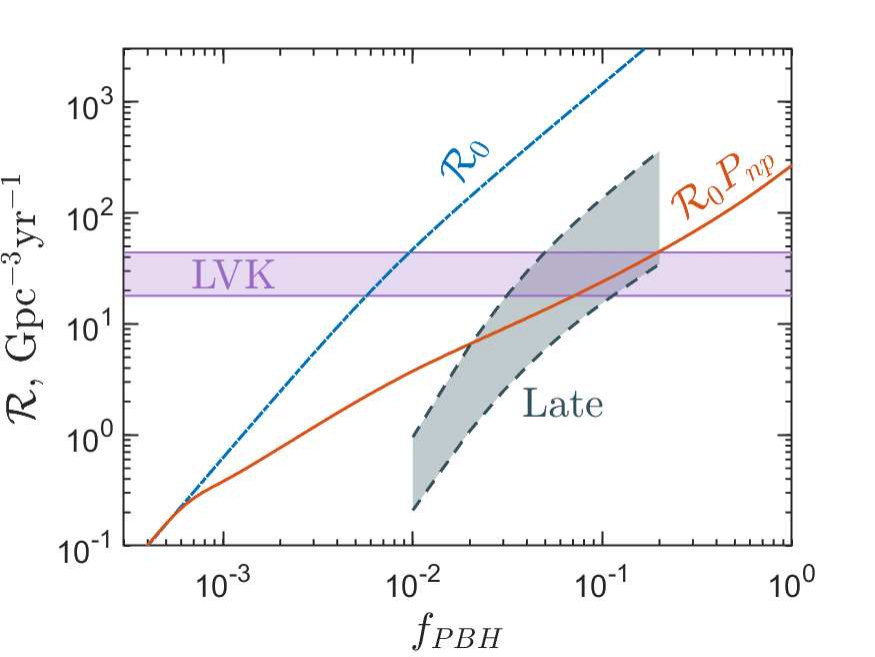}
	\end{center}
\caption{The PBH merger rate at $z = 0$ depending on their fraction in the DM composition $\fbh$. The gray area bounded by dashed lines corresponds to late binaries and is given by Eq.\ \eqref{mr_l1}, where the upper assumes $w = 1$, the lower one is calculated according to Section~\ref{sec3}. Horizontal shaded area is the LVK observations $ \mathcal{R} = 17.9 \div 44 $~Gpc$^{ -3}$~yr$^{-1}$ \cite{KAGRA:2021duu}. The solid line corresponds to the merger rate of early binaries taking into account the suppression factor $P_\np$ (mergers of unperturbed binaries), the dot-dashed line is without suppression}
	\label{mr_modern_f} 
\end{figure}

Figure~\ref{mr_late_diff_z_f} shows the dependence of the late binaries merger rate on redshift for different moments of halo formation $z_f$. Here we neglect halo destruction $w = 1$. Gravothermal instability, which leads to an increase in the density of PBHs in the center of the halo, also leads to the fact that the rate of PBHs mergers increases by two orders of magnitude. It can be seen that the modern merger rate will be maximum if the halos are formed at $z_f \approx 20 - 30$. For smaller $z_f$, the halos turn out to be less dense, and the most massive of them have not experienced the core collapse by the $t_0$. On the other hand, early formed halos correspondingly transit to the expansion stage earlier, so the PBHs merger rate in them is lower due to their lower density. The modern merger rate of late binaries weakly depends on the specific moment of halo formation $z_f$ and the difference is only a few times. Nevertheless, for a rough estimate of the merger rate we will take the maximum of the expression in Eq.~\eqref{mr_l1} as a function of $z_f$.

Figure~\ref{mr_modern_f} shows the modern rate of PBH mergers depending on their fraction in the composition of dark matter $\fbh$. The gray-green area bounded by dashed lines is the merger rate of late binaries. The upper dashed line corresponds to $w = 1$ in Eq.~\eqref{mr_l1}; i.e., halos are not destroyed. While the bottom line corresponds only to the surviving halos and their fraction $w$ is calculated according to Section~\ref{sec3} and shown in Fig.~\ref{surv_frac}. We emphasize that both cases are obtained as a result of maximizing the merger rate in Eq.~\eqref{mr_l1} by $z_f$. The merger rate of unperturbed early binaries is shown by the red solid line. It can be seen that early binaries can have a subdominant contribution to PBHs mergers. In particular, as seen, taking into account the suppression factor, the merger rate of early binaries does not contradict the LVK observations for the PBHs fraction $\fbh = 0.1$. Mergers of late binaries also satisfies the LVK observations, but at the lower limit. If we do not take into account the destruction of clusters, then it will exceed the LVK observations several times. On the other hand, for $\fbh < 0.05$, the merger rate of both late and early binaries does not contradict observations. 

The formation of clusters in which PBHs actively interact with each other inevitably leads to a decrease mergers of early binaries. At the same time, the probability of late binaries mergers increases in clusters. Therefore, the red curve and shaded area in Fig.~\ref{mr_modern_f} are related. Additionally, the red line implies maximum suppression of early binaries mergers. However, perturbing binaries is more effective at high redshifts, because the halo has time to experience core collapse before being absorbed into a large halo, see Fig.~\ref{P_ab}. For example, at $z = 10$, the fraction of unperturbed binaries from Fig.~\ref{Sz_fpbh} is $P_{\rm np} \approx 0.04$, which corresponds to the moment of formation of $z_f \approx 30$. 

\section{High-redshift PBH mergers}\label{sec5}

The uniqueness of PBHs from the point of view of observing gravitational wave signals is that their merger rate monotonically increases with redshift. Therefore, the observation of black hole mergers at high redshifts will unambiguously answer the question about the origin of these black holes: astrophysical or primordial. In the absence of clustering effects, the merger rate of early binaries changes over time as $\mathcal{R} \propto (t(z) / t_0)^{-34/37}$~\cite{Vaskonen:2019jpv}. When taking into account clustering, the growing nature of the dependence of $\mathcal{R}$ on time is preserved. However, as was shown in the previous section, mergers of late PBH binaries forming in clusters can exceed the merger rate of early binaries in the modern era. Figure~\ref{mr_late_diff_z_f} shows that mergers of late binaries make a significant contribution only at relatively low redshifts $z \lesssim 5$. In this region of redshifts, it is difficult to predict the form of the $\mathcal{R}(z)$ dependence, because it is necessary to take into account the mergers of both channels for formation of PBH binaries. In addition, it is necessary to take into account mergers of black holes of stellar origin. However, at high redshifts $z > 5$, early binaries make a dominant contribution to the rate of black hole mergers $\mathcal{R}(z) = \mathcal{R}_0(z) P_{\rm np}(z)$, which is shown by the dashed black line in Fig.~\ref{mr_late_diff_z_f}.

\begin{figure}
\centering
\includegraphics[width=0.5\textwidth]{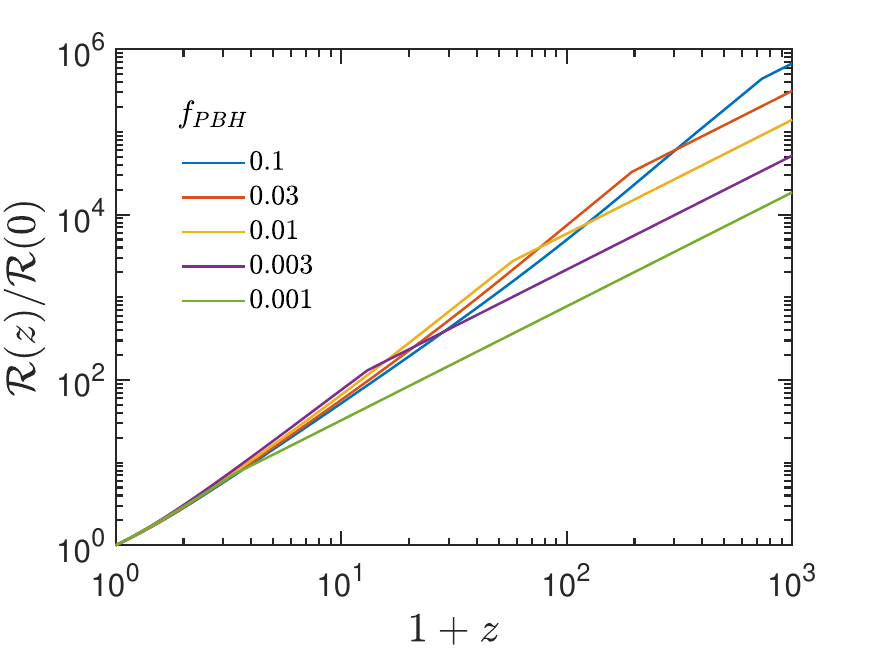}
\includegraphics[width=0.5\textwidth]{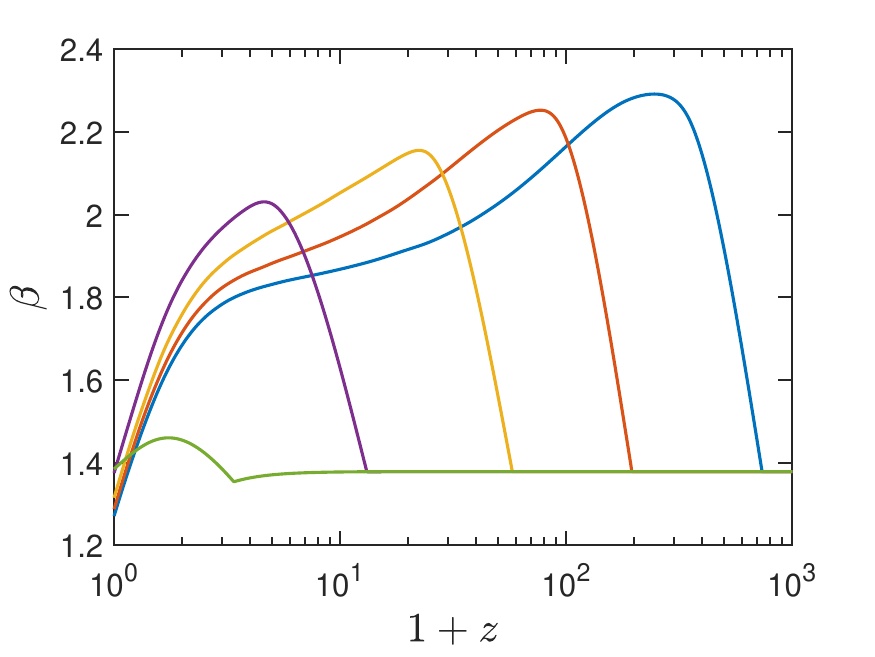}
\caption{Top: the dependence of normalized merger rate of early binaries from redshift for different fraction of PBH in the DM $\fbh$
Bottom: the logarithmic slope $\beta = d \ln \mathcal{R} / d \ln(1+z)$ the color of the lines is the same as in the legend in the top panel}
\label{mr_slope}
\end{figure} 

The top panel of Fig.~\ref{mr_slope} shows the redshift dependence of the merger rate of early PBH binaries. The bottom panel shows the logarithmic slope $\beta$ of the merger rate $\mathcal{R} \propto (1 + z)^{\beta}$. It can be seen that at high redshifts the clustering effects become negligible and the slope changes to $\beta = 1.4$, which corresponds to all early binaries being unperturbed. If the PBH fraction is $\fbh \sim 0.1$, then the merger rate scales as $\mathcal{R} \propto (1 + z)^{1.9}$. With such a fraction of PBHs in the DM composition, clustering effects turn out to be important up to the moment of recombination $z \approx 10^3$. On the other hand, for $\fbh \sim 0.01$ the change of exponent $\beta$ to the case of completely unperturbed binaries $\beta = 1.4$ occurs at redshifts $z \lesssim 50$, which is potential observable by future generation gravitational wave detectors~\cite{Franciolini:2023opt, Branchesi:2023mws}. 

In the case of a small fraction of PBHs $\fbh \lesssim 10^{-3}$, the merger rate changes as $\mathcal{R} \propto (1 + z)^{1.4}$, because there are no clustering effects. We neglect the possible perturbation of binaries in standard dark halos arising from inflationary fluctuations. For a small fraction of $\fbh$, the main contribution to mergers is due to black holes of stellar origin. Their merger rate is related to the star formation rate and is at maximum $\mathcal{R} \simeq 100$~Gpc$^{-3}$~yr$^{-1}$ at redshifts $z\sim 1 - 2$~\cite{Mandic:2016lcn, Sasaki:2018dmp}. An additional peak of smaller height is also possible at redshifts $z \approx 10$ from remants of Population III stars~\cite{Ng:2020qpk, Ng:2022agi}. At even higher redshifts, binary black holes are difficult to produce through astrophysical channels, so their merger rate decreases rapidly. The conservative threshold for mergers of such black holes corresponds to $z \sim 30$~\cite{Koushiappas:2017kqm}. At that time, the rate of PBH mergers monotonically increases with redshift for any values of $\fbh$. Therefore, the observation of black hole mergers at redshifts $z \gtrsim 30$ will indicate their primordial origin. 

From the point of view of high redshifts observations, the number of gravitational wave events per unit time that occur in a certain range of redshifts is of interest~\cite{Nakamura:2016hna, DeLuca:2021hde, Martinelli:2022elq}:
\begin{equation} \label{Nev}
    N_{\rm events} = \int_{z_{\rm min}}^{z_{\rm max}} dz \, \frac{\mathcal{R}(z)}{1 + z} \frac{d V_c}{dz}.
\end{equation}
Here the factor $1 + z$ in the denominator converts the rate in the source frame to the observer's frame and $dV_c / dz$ is the comoving volume of the spherical shell between $z$ and $z + dz$ 
\begin{equation}
    \frac{dV_c}{dz} = 4 \pi r^2(z) \frac{c}{H(z)}
\end{equation}
$r(z)$ is the comoving distance at redshift $z$
\begin{equation}
    r(z) = \int_0^{z} dz' \frac{c}{H(z')}.
\end{equation}
We use $H(z) = H_0 \sqrt{\Omega_{\Lambda} + \Omega_M(1+z)^3}$ with $H_0 = 67.4$~km~s$^{-1}$~Mpc$^{-1}$, $\Omega_M = 0.315$ and $\Omega_{\Lambda} = 0.685$~\cite{Planck:2018vyg}. In the integral in Eq.\ \eqref{Nev} we mean mergers of early binaries, since for them it is easy to obtain the merger rate on time dependence. Also, this integral is a weakly dependent function of the upper limit, we set it equal to $z_{\rm max} = 50$. The lower limit is $z_{\rm min} = 5$, since late binary mergers become significant at lower redshifts. Note that Eq.\ \eqref{Nev} corresponds to the number of signals registered by an ideal detector. A real detector, in turn, has some efficiency, so the observed number of gravitational wave events will be significantly lower. 

\begin{figure}
	\begin{center}
\includegraphics[angle=0,width=0.5\textwidth]{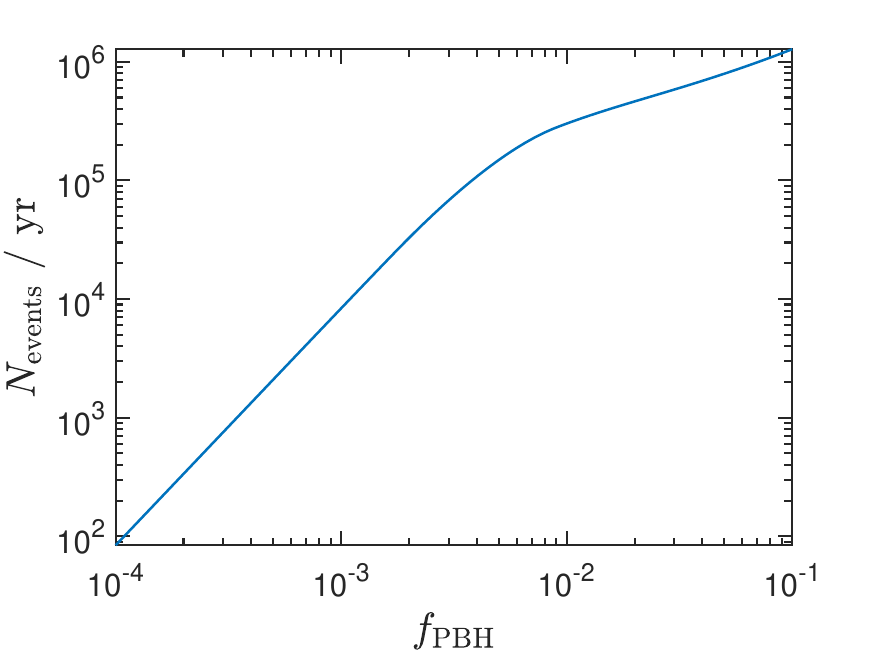}
	\end{center}
\caption{Number of PBH mergers per year, at redshifts $5<z<50$ depending on $\fbh$ is given by Eq.~\eqref{Nev}}
	\label{Nev_gr} 
\end{figure}

Figure~\ref{Nev_gr} shows the number of PBH mergers per year at high redshifts $z > 5$ depending on $\fbh$. We assume that $\fbh \leq 0.1$ since this does not contradict the gravitational wave constraints on the observed merger rate, see Fig.~\ref{mr_modern_f}. At $\fbh > 0.01$, clustering effects are relevant even at redshifts $z > 100$, which leads to active perturbations of early binaries. It can be seen that up to $\mathcal{O}(10^6)$ black hole merger events per year can be expected, occurring at high redshifts. 

\section{Discussion} \label{sec6}

In this work, the evolution of early DM halos formed as a result of the initial Poisson distribution of PBHs was studied. The internal dynamics of these halos is determined by the gravitational interaction of PBHs both with each other and with dark matter particles. Similar to how it is done in astrophysical applications to study the evolution of globular star clusters, we used the Fokker-Planck kinetic equation. Due to gravothermal instability, the evolution of dark halos leads to the formation of dense PBH clusters surrounded by a massive DM halo. Moreover, in the case when PBHs of tens solar masses are a subdominant component of DM $\fbh \lesssim 0.1$, the timescales for the formation of a cluster in the halo turn out to be comparable with the characteristic time of structure formation --- Hubble time. Roughly speaking, the cluster forms before being absorbed into a large halo. In the opposite case $\fbh \approx 1$ this is no longer mostly the case. The time required for cluster formation (core collapse time) significantly exceeds the time of structure formation, we discuss in detail this in Section~\ref{sec3}. Therefore, PBHs with a narrow mass distribution centered at $\sim 10 \, M_{\odot}$ as the dominant component of dark matter are excluded by observations of gravitational waves. It has also been shown that emerging PBH clusters can survive the process of structure formation if they are absorbed by large halos at redshifts $z < 4 $. We estimate that about $10 - 40\%$ clusters can survive to the modern epoch. 

We also considered the influence of PBH clustering on the perturbation of the parameters of PBH binaries forming in the early Universe. It was shown that a large fraction $\gtrsim 0.95$ of such binaries are perturbed in early structures, that leads to a strong decreasing of the PBH merger rate. Similar to the formation of clusters, the processes of perturbation of early binaries occur more efficiently for the case $\fbh < 1$. We also considered the issue of merging late binaries of PBHs in clusters. The parameters of the forming binaries are predominantly such that the binaries immediately merge after formation. In addition, the internal evolution of the halo leads to the fact that this channel may be dominant for PBHs mergers in the current epoch. It is important to emphasize that the dominant character in late binaries mergers and the perturbations of early binaries turn out to be related: the suppression of the merger rate of early binaries inevitably leads to an increase in mergers of late binaries. Since both effects require the formation of clusters.  

Late binaries are important for PBH mergers only in a relatively modern era, however it depends on the destruction of clusters during the structure formation. If most of the clusters are destroyed, then the contribution of late binaries to the PBH merger rate is subdominant. If the clusters survive completely, then the situation is reversed and PBH mergers dominate in them up to redshifts $z \lesssim 5$. In this redshift region, it is difficult to predict the behavior of the merger rate, because the contribution of early and late binaries is comparable. In addition, as the redshift decreases, the fraction of surviving clusters will also decrease. For this reason, the rate of late binaries mergers shown in Fig.~\ref{mr_late_diff_z_f} probably becomes more reliable as redshifts increase. 

In this work, we studied the redshift evolution of early binaries merger rate $\mathcal{R} \propto (1 + z)^{\beta}$. If PBH binaries are not perturbed, then the exponent $\beta = 1.4$. This situation is realized in two cases: a) if $\fbh \lesssim 10^{-3}$, in this case there is no clustering in principle; b) at high redshifts, where Poisson clustering has not yet begun. The latter case can be seen in Fig.~\ref{mr_slope}, when the $\beta$ drops sharply to $1.4$. In particular, if the fraction of PBHs is $10^{-3} < \fbh < 10^{-2}$, then the transition to the region of the absence clustering occurs at $z \lesssim 50$. For the case $\fbh > 0.01$, the influence of clustering on the mergers of early binaries is significant at all redshifts. The exponent of the merger rate can reach $\beta \approx 2.2$. The obtained PBH merger rate can be verified using third-generation gravitational wave detectors Einstein Telescope and Cosmic Explorer, which will be able to observe black hole mergers of tens solar masses up to redshifts $z \lesssim 100$. 

In this study, the main source of gravitational wave events at high redshifts are mergers of PBHs that form before matter-radiation equality and escape Poisson clustering. However, if the observed merger rate at high redshifts differs from the analysis presented in this paper, then this may indicate an initially clustered distribution of PBHs. Also, binaries that did not fall into Poisson clusters can merge in earlier epochs due to interaction with dark matter particles \cite{Hayasaki:2009ug}. In these cases, the main contribution to mergers will be due to late binaries in clusters. 

Our analysis also considers the monochromatic mass spectrum of PBHs $m = 30 \, \Msun$. Taking into account the wide mass distribution will have little effect on our results. If PBH with masses $1 - 100\,\Msun$ are a subdominant part of dark matter, then the dynamics of early halos is also driven by dynamical friction against dark matter particles. Therefore, the timescales for the perturbations of early binaries and the formation of dense primordial black hole clusters will qualitatively correspond to our results. Also, recent reanalysis of the microlensing constraints in Ref.\ \cite{Garcia-Bellido:2024yaz} indicates that all dark matter can be explained by PBHs (although Ref.~\cite{Mroz:2024mse} argues that such possibility is excluded). In this case, the peak of the mass distribution of PBHs corresponds to solar masses~\cite{Carr:2019kxo}. In this scenario the dynamics of PBHs of tens solar mass (including binaries) should also generally agree with our results. 

\section*{Acknowledgement} 

I am grateful to Konstantin Belotsky for useful discussions and critical remarks. I also thank the anonymous referee for the very constructive comments, which helped to improve the paper. The work was supported by RSF grant \textnumero 23-42-00066. 

\section*{Appendix: Fokker Planck equation} 

The one-dimensional orbit averaged Fokker-Planck equation describes the evolution of the distribution function $f(E)$ due to diffusion in energy space~\cite{1980ApJ...242..765C, 1987ApJ...319..801L} 
\begin{equation} \label{fp_Eq}
    \frac{\partial N_i}{\partial t} = - \frac{\partial}{\partial E} \left ( m_i D_E f_i + D_{E,b}f_i + D_{EE} \frac{\partial f_i}{\partial E} \right),
\end{equation}
here $N_i(E) = 4 \pi^2 p(E) f_i(E)$ is the number density in the energy space of particles $i$-th type (by $i$ we mean either PBHs or DM particles). The factor $p(E)$ is written in the form 
\begin{equation}
    p(E) = 4 \int_{0}^{\phi^{-1}(E)} d{r} \, r^2 \sqrt{2 \big( E - \phi(r) \big)}, 
\end{equation}
where $\phi(r)$ is the gravitational potential. In the spherically symmetric case, the Poisson equation $\nabla^2 \phi = 4 \pi G \rho$ has a solution
\begin{equation} \label{gr_pot}
    \phi(r) = -4 \pi G \left ( \frac{1}{r} \int_0^r \, d{r'} \, r'^2  \rho(r') + \int_r^{\infty} d{r'} \, r' \rho(r')  \right),
\end{equation}
where the density profile is related to the distribution function according to
\begin{equation} 
    \label{rho}
    \rho(r) = 4 \pi \sum_{j} m_j \int_{\phi(r)}^{0} dE \, f_j(E) \sqrt{2 \big( E - \phi(r) \big)}.
\end{equation}

The coefficient $D_{E,b} = -4 \pi^2 p(E) \langle \Delta E \rangle_{b}$ in Eq.~\eqref{fp_Eq} is responsible for three-body binary heating and $\langle \Delta E \rangle_b$ is the expression averaged along the orbit in Eq.~\eqref{E_bheat}
\begin{equation}
    \langle \Delta E \rangle_{b} = \frac{\displaystyle \int_0^{\phi^{-1}(E)} dr \, r^2 \dot{E}_{b} \sqrt{ \phi(r) - E}}{\displaystyle \int_0^{\phi^{-1}(E)} dr \, r^2 \sqrt{ \phi(r) - E}}.
\end{equation}
The coefficients $D_{E}$ and $D_{EE}$ in Eq.~\eqref{fp_Eq} are given by the following expressions 
\begin{align}
    D_{E} &= -16 \pi^3 \Gamma \sum_j m_j \int_{\phi (0)}^E dE' \, f_j(E') p (E'), \nonumber \\
    D_{EE} &= - 16 \pi^3 \Gamma \sum_j m_j^2 \left [ q(E) \int_E^0 dE' \, f_j(E') \right. \nonumber \\
    &\left.+ \int_{\phi(0)}^{E} dE' \, q(E') f_j(E') \right]
\end{align}
here $\Gamma = 4 \pi G^2 \ln{\Lambda}$ with $\ln{\Lambda}$ is the Coulomb logarithm and $q(E)$ is proportional to the volume of the phase space for particles with energies $\leq E$
\begin{equation} \label{qE}
    q(E) = \frac{4}{3} \int_{0}^{\phi^{-1}(E)} dr \, r^2 \Big[ 2 \big( E - \phi(r) \big) \Big]^{3/2},
\end{equation}
we note that $q(E)$ is the adiabatic invariant.

To solve the Eqs.~\eqref{fp_Eq} and~\eqref{gr_pot} together, we first follow the work~\cite{2017ApJ...848...10V} and use the variable $q$ instead of $E$ and go to the logarimic grid $x = \ln q$. The advantage of using $q$ as an independent variable will become clearer later. The solution itself is divided into a Fokker-Planck step and a Poisson step. The first step is that for time $\Delta t$ the distribution function $f(q)$ is advanced through solving the Fokker-Planck equation using the finite-difference Chang-Cooper scheme~\cite{1970JCoPh...6....1C}. The second step is to update the gravitational potential so that the Eqs.~\eqref{gr_pot} and~\eqref{rho} are consistent~\cite{1979ApJ...234.1036C}. Using the new distribution function and the gravitational potential from the previous time step, the density is found 
\begin{equation} \label{eq.rho_phi}
    \rho (\phi^{\rm old}) = 4 \pi m \int_{\phi^{\rm old}}^{0} dE \, f^{\rm new} \sqrt{2 \big( E - \phi^{\rm old} \big)},
\end{equation}
where for simplicity we have omitted the sum sign in Eq.~\eqref{rho}. Next, the Poisson equation is solved by iterations $\phi^{n+1} = \mathcal{L} \rho (\phi^n)$ until the required accuracy is achieved, where the operator $\mathcal{L}$ is given by Eq.~\eqref{gr_pot}. Since $q$ is the adiabatic invariant, the distribution function $f(q)$ remains fixed when the gravitational potential changes. In order to integrate Eq.~\eqref{eq.rho_phi}, it is necessary to construct the mapping between the values of $q$ and $E$ using Eq.~\eqref{qE}.

\bibliography{bib.bib}

\end{document}